\documentclass[prd,twocolumn,showpacs,preprintnumbers,amsmath,amssymb,superscriptaddress]{revtex4}
\usepackage{graphicx}
\usepackage{epsfig}
\usepackage{rotating}
\usepackage{dcolumn}
\usepackage{bm}
\newcommand{\lsim}{\raisebox{-0.13cm}{~\shortstack{$<$ \\[-0.07cm] $\sim$}}~}
\newcommand{\gsim}{\raisebox{-0.13cm}{~\shortstack{$>$ \\[-0.07cm] $\sim$}}~}
\def\beq{\begin{equation}}
\def\eeq{\end{equation}}

\def\bea{\begin{eqnarray}}\def\eea{\end{eqnarray}}
\def\slash{\!\!\!/}
\begin{document}
\preprint{{FTUV-11-12-20, IFIC/11-71, CERN-PH-TH/2011-313}}
\title{Non-perturbative effects in WIMP scattering off nuclei
in the NMSSM}
\author{Grigoris Panotopoulos}
\affiliation{CERN, Theory Division, CH-1211 Geneva 23, Switzerland}
\author{Miguel-Angel Sanchis-Lozano}
\affiliation{Instituto de F\'{\i}sica Corpuscular (IFIC) and
Departamento de F\'{\i}sica Te\'orica, Centro Mixto Universitat de
Val\`encia-CSIC, Dr. Moliner 50, E-46100 Burjassot, Valencia, Spain}

\begin{abstract}
We explore a scenario in the Next-to-Minimal-Supersymmetric-Standard-Model
(NMSSM) with both a light ${\cal O}(10)$ GeV
neutralino and a CP-odd Higgs boson
with significant coupling to down-type fermions,
evading all current B physics, LEP and WMAP bounds.
Motivated by a possible slight
lepton universality breaking hinted in $\Upsilon$ decays,
we consider the effect of the mixing
of $\eta_b$ resonances with the pseudoscalar Higgs
on the spin-dependent scattering neutralino cross section
off nucleons.
We conclude that this mechanism could be relevant 
provided that
non-perturbative effects enhance the effective $\eta_b$-nucleon
coupling, taking over velocity/$q^2$ suppression factors, 
perhaps giving a new insight into the current controversial
situation concerning direct search experiments of dark matter.
\end{abstract}
\pacs{12.60.Fr, 12.60.Jv, 13.20.Gd, 95.35.+d, 98.80.Cq, 98.80.Es}
\maketitle

\section{Introduction}
Evidence has been accumulated both from astrophysics and cosmology
that about 1/4 of the energy budget of the present universe consists
of the so-called (cold) dark matter (DM), namely, a component which
is non-relativistic and neither feels the electromagnetic nor the
the strong interaction. It is fair to say that the most popular DM
candidate for a WIMP (Weakly Interacting Massive Particle) is the
lightest supersymmetric particle (LSP) in supersymmetric models with
$R$-parity conservation. Leaving aside the axion and the axino, the
superpartners with the right properties for playing the role of
a WIMP in the universe are the gravitino and the lightest
neutralino ($\chi$) - by far, the most discussed case in the literature.

Although the LHC is running smoothly and collecting large amounts of
data useful to look for physics beyond the Standard Model (SM),
other complementary facilities are certainly needed, especially
concerning DM detection. In fact, DAMA/LIBRA, CoGeNT, and
more recently CRESST experiments have
reported the observation of events in excess of the expected
background, hinting at the existence of a light
WIMP~\cite{Bernabei:2010mq,Aalseth:2011wp,Angloher:2011uu}.
However, exclusion limits
set by other direct searches, such as Xenon10~\cite{arXiv:1104.3088}  
and Xenon100~\cite{arXiv:1104.2549}, 
are in tension with the above claims.

In the NMSSM, a light neutralino (as a DM candidate) can efficiently
annihilate through the resonant s-channel via a light pseudoscalar 
Higgs mediator satisfying the requirements from the relic 
density~\cite{Gunion:2005rw}.
However, following a scan of the NMSSM 
parameter space, the authors of
\cite{Das:2010ww} obtained upper limits on the
spin-independent (SI) $\chi$-nucleon cross section which
are substantially below the requirements of DAMA and CoGeNT. 
The spin-dependent (SD) cross section (via $Z$-exchange) 
was also found several
orders of magnitude below current experimental bounds. 
On the other hand,  
the authors of \cite{Gunion:2010dy} were able to 
achieve a somewhat larger SI cross section
in a similar scenario. Admittedly, such cross
sections can be further enhanced by
increasing the $s$-quark content of the nucleon, but the
agreement with the low range of DAMA results turns out to be
only marginally acceptable.

In this work we revisit
SD $\chi$-nucleon scattering
via pseudocalar-exchange
in the NMSSM, usually neglected in
most analyses \cite{Das:2010ww}, which however might be enhanced due to
a non-perturbative mechanism as later argued.

\begin{table*}[hbt]
\caption{Phase-space corrected
leptonic branching fractions, $\hat{\cal B}\left(\Upsilon(nS)
\to \ell \ell\right)$ (in \%), and error bars (summed in quadrature) of
$\Upsilon(1S)$, $\Upsilon(2S)$, and $\Upsilon(3S)$ resonances
\cite{pdg}. Error bars of the ratios
$\hat{R}_{\tau/\ell}(nS)$ are likely overestimated
because of expected correlations between the numerator
and denominator experimental uncertainties.}

\label{FACTORES}

\begin{center}
\begin{tabular}{|c|c|c|c|c|c|}
\hline
 & $\hat{\cal B}\left(e^+e^-\right)$ & $\hat{\cal B}\left(\mu^+\mu^-\right)$ &
 $\hat{\cal B}\left(\tau^+\tau^-\right)$ & $\hat{R}_{\tau/e}(nS)$ &
$\hat{R}_{\tau/\mu}(nS)$\\
\hline
$\Upsilon(1S)$ & $2.48 \pm 0.07$ & $2.48 \pm 0.05$ & $2.62 \pm 0.10$ &
${\bf 0.057 \pm 0.050}$ & ${\bf 0.057 \pm 0.046}$ \\ \hline
$\Upsilon(2S)$ & $1.91 \pm 0.16$ &  $1.93 \pm 0.17$ & $2.01 \pm 0.21$ &
${\bf 0.052 \pm 0.141}$ & ${\bf 0.041 \pm 0.141}$ \\ \hline
$\Upsilon(3S)$ & $2.18 \pm 0.21$ & $2.18 \pm 0.21$ & $2.30 \pm 0.30$ &
${\bf 0.056 \pm 0.171}$ & ${\bf 0.056 \pm 0.171}$ \\  \hline \hline
$\psi(2S)$ & $0.773 \pm 0.017$ & $0.77 \pm 0.08$ & $0.772 \pm 0.100$ &
${\bf -0.001 \pm 0.100}$ & ${\bf 0.002 \pm 0.100}$ \\

\hline
\end{tabular}
\end{center}
\end{table*}

\section{A light NMSSM CP-odd Higgs boson}

The Higgs sector of the NMSSM 
contains six independent parameters:
$\lambda,\ \kappa,\ A_{\lambda}, \
A_{\kappa},\ \tan{\beta}$ and $\mu_{eff}$, whose definitions can be found
elsewhere~\cite{arXiv:0910.1785}. Notice that $\mu_{eff}= \lambda s$ is
generated as the vev of the singlet field  $s \equiv
\left<S\right>$; it is also useful to define $B_{eff}=A_\lambda +
\kappa s$.

Two physical pseudoscalar states appear in the spectrum of the NMSSM
as superpositions of the MSSM-like 
state $A_{MSSM}$ and the singlet-like state $A_S$. 
In particular for the lightest CP-odd Higgs boson
\begin{equation}
A_1=\cos{\theta_A}A_{MSSM}+\sin{\theta_A}A_S\; ,
\end{equation}
where $\theta_A$ stands for the  mixing 
angle~\cite{Domingo:2008rr}.  
The $A_1$ reduced coupling $X_d$ 
to down-type quarks and leptons (normalized with respect to the
coupling of the CP-even Higgs boson of the SM) reads
\begin{equation}\label{eq:Xd}
X_d=\cos{\theta_A}\tan{\beta} \simeq -\frac{\lambda v\ 
(A_{\lambda}-2 \kappa s)}
{M_A^2+3 \kappa A_{\kappa}s} \times \tan{\beta}\; ,
\end{equation}
where $M_A^2 = 2 \mu_{eff} B_{eff}/\sin{2\beta}$. 

An analysis of a particular region of
the NMSSM parameter space where 
$X_d$ can be relatively large
at high $\tan{\beta}$, 
together with a light CP-odd Higgs boson ($m_{A_1} \sim {\cal O}(10)$ GeV),
was carried out in \cite{Domingo:2008rr} 
although without including the relic abundance constraint. 
Let us stress here again that this scenario is 
quite different from the
PQ-symmetry-limit ($\kappa \to 0$) or R-symmetry-limit ($A_\kappa,\
A_\lambda \to 0$), where $X_d$ remains moderate even in the
large $\tan{\beta}$ limit. Although not motivated
by any symmetry as in the latter cases, we remark 
that $\tan\beta \sim 1/|\mu_{eff} B_{eff}|$ 
for large values of $\tan\beta$ \cite{Ananthanarayan:1996zv}, 
giving consistency to our scenario
which implies (relatively) small values of $|B_{eff}|$. 

As pointed out in Refs.~\cite{Fullana:2007uq,Domingo:2009tb} a
large $X_d$ could induce a non-negligible mixing of the 
CP-odd state and $\eta_b(nS)$ hadronic resonances. For the sake 
of simplicity, 
we only consider the $A_1$ mixing
with the nearest (in mass) pseudoscalar resonance, generically denoted
hereafter as $\eta_{b0}$. 
Thus, the $A_1$ and $\eta_b$ physical states can be 
written approximately as
\begin{eqnarray}\label{eq:mixing1}
A_1 &=& \cos{\alpha}\ A_{10}\ +\ \sin{\alpha}\ \eta_{b0} \\
\eta_b &=& \cos{\alpha}\ \eta_{b0}\ -\ \sin{\alpha}\ A_{10}
\label{eq:mixing2} 
\end{eqnarray}
where subindex zero refers to
unmixed states throughout. (The dominant components
may of course be reversed if $\alpha > \pi/4$.) In any event,   
one should keep in mind that 
the $A_{10}$ can mix (to a greater or lesser extent) with more 
than a single pseudoscalar hadronic state (see Ref.~\cite{Domingo:2009tb}).

The strength of the mixing is determined by
the angle $\alpha$ given by~\cite{Drees:1989du,Domingo:2008rr}
\beq\label{eq:alpha}
\sin{}2\alpha\ =\ \biggl[1+\frac{(m_{A_{0}}^2-m_{\eta_{b0}}^2)^2}{4\ 
\delta m^4}\biggr]^{-1/2}\; ,
\eeq 
where the imaginary part has been neglected and $\delta m^2$ can be computed by
means of a non-relativistic quark potential model:
\begin{equation}
\delta m^2\ =\
\biggl(\frac{3m_{\eta_{b}}^3}{4 \pi v^2}\biggr)^{1/2} |R_{\eta_{b0}}(0)|
\times X_d\; ,
\end{equation}
where $v=246$ GeV and $R_{\eta_{b0}}(0)$ stands for the 
radial wave function at the origin of the 
corresponding $\eta_{b0}$ state (for
more details see Ref.~\cite{Domingo:2008rr}).

As emphasized in Ref.~\cite{SanchisLozano:2010ij}, a substantial mixing
of a ${\cal O}(10)$ GeV CP-odd Higgs boson with $\eta_{b0}$ resonances
can modify (hindering) a signal based on direct observation of  
a monochromatic peak in the photon spectrum
of radiative $\Upsilon$ decays \cite{hep-ph/0612031}. 
On the other hand, 
a light CP-odd Higgs could still show up as a slight
breaking of lepton universality in 
the ratio ${\cal B}_{\tau\tau}/{\cal
B}_{\ell\ell} \approx 1$, where ${\cal B}_{\tau\tau}$ denotes the tauonic, and
${\cal B}_{\ell\ell}$ the electronic ($\ell=e$) or muonic
($\ell=\mu$) branching ratios of the $\Upsilon$ resonance,
respectively \cite{Sanchis-Lozano:2003,SanchisLozano:2006gx}.

In view of the greatly improved accuracy of the recent 
measurements of the leptonic BFs 
(and likely so in the forthcoming BaBar analysis of 
$\Upsilon(3S)$ decays \cite{Roney}),
it seems advisable to remove the dependence
on the final-state lepton mass ($m_{\ell}$) dividing the
branching fraction (BF) by $K(x_{\ell})=(1+2x_{\ell})(1-4x_{\ell})^{1/2}$,
which behaves as a (smoothly) decreasing function of
$x_{\ell}=m_{\ell}^2/M_{\Upsilon}^2$:
${\hat{\cal B}_{\ell\ell}}={\cal B}_{\ell\ell}/K(x_{\ell})$, with
$\ell=e,\mu,\tau$. Therefore defining
\begin{equation}
\hat{\cal R}_{\tau/\ell}= \frac{\hat{\cal B}_{\tau\tau}-
\hat{\cal B}_{\ell\ell}}{\hat{\cal B}_{\ell\ell}}=
\frac{\hat{\cal B}_{\tau\tau}}{\hat{\cal B}_{\ell\ell}}-1\ \ ;\ \
\ell=e,\mu\; ,
\label{eq:Rhat}
\end{equation}
the contribution of a pseudoscalar Higgs to the (inclusive)
decay rate 
would imply an enhancement
of the tauonic mode and therefore small but positive values of
$\hat{\cal R}_{\tau/\ell}$ \cite{Sanchis-Lozano:2003}. 

The experimental results obtained from
the PDG listing \cite{pdg} are shown in Table~1. The good agreement
of the $\psi(2S)$ measurements with the SM expectations, 
together with the systematic
disagreement of the $\Upsilon$ family, are consistent with a slight
enhancement of the tauonic decay mode of $\Upsilon$ resonances 
versus the electronic and muonic decay modes, due
to an extra contribution of
a light pseudoscalar Higgs boson, which couples to
down-type quarks (at large $\tan{\beta}$) but of negligible effect
for up-type quarks in this limit.

Furthermore, 
unexpected values for the hyperfine splittings in the bottomonium
spectrum: $\Delta E_{hyp}(nS)=m_{\Upsilon(nS)}-m_{\eta_b(nS)}$
($n=1,2,3$)~\cite{Fullana:2007uq,Domingo:2009tb} can
be induced by the mixing (\ref{eq:mixing1}-\ref{eq:mixing2}). In particular,
$\Delta E_{hyp}(1S)= 69.3 \pm 2.8$ MeV obtained by BaBar and CLEO using
hindered radiative $\Upsilon(3S)$ decays \cite{pdg}
appears to be significantly
larger than expected from perturbative QCD, estimated to be $42 \pm
13$ MeV  \cite{Kniehl:2003ap}. However,  
the recent Belle measurement
of the $\eta_b(1S)$ mass based on the $h(1P) \to \eta_b(1S) \gamma$ decay 
\cite{Collaboration:2011chb}
leads to $\Delta E_{hyp}(1S)= 59.3 \pm 3.1$ MeV, in very good agreement
with lattice NRQCD calculations \cite{Meinel:2010pv}. 
Note in passing that the discovery of the
$\eta_b(2S)$ state and its mass determination would provide a crucial
check when compared to the lattice
prediction for $\Delta E_{hyp}(2S)$ \cite{Meinel:2010pv}.

\subsection{NMSSM scan including WMAP bounds}

In order to assess the impact of the 
$A_{10}-\eta_{b0}$ mixing on 
DM phenomenology, we first have to
review the present bounds from B physics, LEP and
DM relic abundance. To this aim the latest version of
NMSSMTools \cite{nmssmtools} was employed 
to scan the NMSSM parameter space
with micrOMEGAs turned on
in the main code. We focus on a
narrow mass window for $A_1$, 
where current
experimental constraints still permit large $X_d$ values
\cite{Domingo:2010am} under examination in this paper.

The following conditions were required to be satisfied:
\begin{itemize}
\item[{\it i)}]
$10 \lsim m_{A_1} \lsim 10.58$~GeV (i.e. below $B\bar{B}$ threshold).
\item[{\it ii)}] Relatively large values
of $X_d$, for large $\tan\beta$.
\item[{\em iii)}] A light neutralino of mass ${\cal O}(10)$ GeV, as a WIMP
candidate satisfying the WMAP bounds.
\end{itemize}

\begin{figure}[ht!]
\begin{center}
\includegraphics*[width=1.\linewidth]{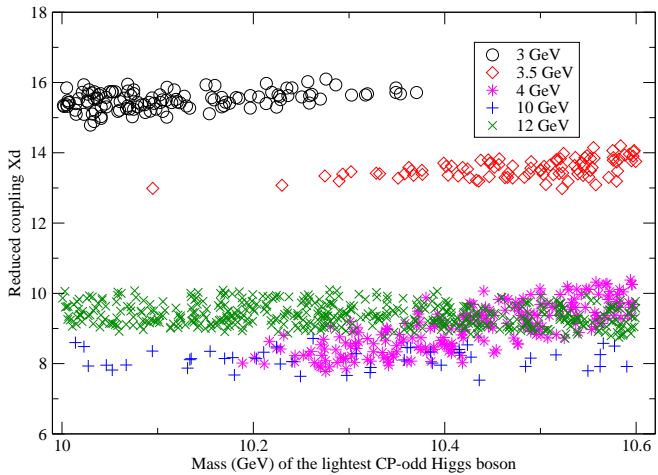}
\end{center}
\caption{$X_d$ versus the mass of the CP-odd Higgs boson ($m_{A_1}$)
obtained from our scans on a particular region of the NMSSM parameter space
for a neutralino mass of a) $\simeq 3$ GeV (black circle)
b) $\simeq 3.5$ GeV (red diamond); c) $\simeq 4$ GeV (magenta asterisk)
d) $\simeq 10$ GeV (blue plus); e) $\simeq 12$ GeV (green cross).}
\label{fig:1}
\end{figure}

The selected ranges of NMSSM parameters
correspond to a particular but
motivated scenario~\cite{Fullana:2007uq,Domingo:2009tb}.
We set $\mu_{eff} \simeq 200$ GeV and $\tan{\beta}=45$ throughout our
analysis, while $\lambda$
and $\kappa$ run over the range $[0.3,0.5]$, with 
$M_A \in [400,550]$ GeV. In order to get the 
highest possible $X_d$
values compatible with the bounds, 
we set $A_{\kappa} \in [-20,-30]$ GeV for $m_{\chi} \simeq 2-3$ GeV,
dropping to $A_{\kappa} \sim -10$ GeV for $m_{\chi} \simeq 10-12$ GeV.
Note that decreasing $|A_{\kappa}|$ implies lowering
$X_d$ as can be inferred from Eq.(\ref{eq:Xd})
keeping $M_A$ fixed within the above interval.

In Fig.~1 we plot $X_d$ versus $m_{A_1}$ for different
values of the neutralino mass, namely $m_{\chi}=$ 3, 3.5, 4, 10 and 12 GeV.
For $m_{\chi}$ close to 5 GeV the values of $X_d$ are
very small as expected on the grounds of a too efficient annihilation
rate of dark matter 
due to the resonant condition: $2m_{\chi} \simeq m_{A_1}$. Notice that
values of $X_d \gsim 8$ are
possible for $m_{\chi} \gsim 10$ GeV, being particularly large
($X_d \gsim 10$) for $m_{\chi} \lsim 4$ GeV. 
It is also worth 
noting that if 
condition {\em iii)} is removed from the scan, the allowed values
of $X_d$ become higher, notably about $m_{\chi} = 5$ since the 
aforementioned 
resonant condition does not apply anymore. 

Let us also remark that the current limit from B factories \cite{2008st} 
on $B[\Upsilon(3S) \to \gamma + \mathrm{invisible}]<(0.7-30) \times 10^{-6}$, 
for $s^{1/2}_{\mathrm{inv}} < 7.8\ \mathrm{GeV}$ where 
$s_{\mathrm{inv}}$ denotes the
missing invariant-mass squared, actually
does not impose any bound on the $\chi$ mass 
(as $m_{A_1} \gsim$ 10 GeV).

\section{Neutralino scattering off nuclei}

Next we address neutralino elastic scattering 
off nuclei based on our analysis
of the NMSSM parameter space.
As is well-known, the SI scattering cross section   
is enhanced by a coherent factor proportional
to the atomic number squared $A^2$. For SD
interactions, the cross section depends on the total spin
of the nucleus and is typically a factor $A^2$ smaller.

The total WIMP-nucleus cross section has
contributions from both the SI and SD interactions, though
one contribution is expected to 
dominate the other depending on the
target nucleus (e.g. according to the even/odd  
number of protons and neutrons) and 
the detection technique employed in the experiment.
The contributions to the SI cross section arise 
in the interaction Lagrangian of the WIMP
with quarks and gluons of the nucleon from
scalar and vector couplings whereas the SD part
is attributed to the axial-vector couplings.
Pseudoscalar interaction is usually neglected because
of a strong velocity and/or momentum transfer suppression.

Nevertheless, momentum-dependent interactions have been put forward~
\cite{astro-ph/0408346,Chang:2009yt,Freytsis:2010ne} in order to
alleviate the tension between the DAMA signal and the null results
from other experiments. In this work we propose 
that a significant $A_{10}-\eta_{b0}$ mixing could dramatically
modify the SD $\chi$-nucleon cross section, in
analogy with the well-known {\em vector-meson-dominance model} (VMD)
for electron (or real photon)
scattering off nuclei \cite{PRINT-77-0549 (HARVARD)},
where the virtual (or real) photon interacts with nucleons
via one of its hadronic components.
In Fig.~2 we depict two graphs illustrating how the mixing
of a pseudoscalar (left), or a photonic mediator (right), 
with hadronic resonances can modify the
effective coupling to the nucleon for WIMP and
electron scattering, respectively.

Likewise, a similar mechanism could be envisaged for the
mixing of scalar resonances (e.g. $\chi_{b0}$ states) and a light
enough CP-even Higgs boson. However, it is widely accepted that,
contrary to the CP-odd Higgs, present bounds
exclude a ${\cal O}(10)$ GeV scalar boson with
relatively large couplings to quarks and leptons (see however
\cite{Draper:2010ew}). In fact, 
the lightest scalar Higgs state in our NMSSM scan has a mass 
$\gsim 110$~GeV, whereby the mixing with hadronic states would be negligible.

The axial-vector and pseudoscalar $\eta_{b0} NN$ couplings defined via
\begin{equation}\label{etaNN}
{\cal L}_{\eta_{b0NN}}=
\frac{g_{\eta_{b0}NN}}{2M_N}\bar{N} q{\slash}\gamma_5N\eta_{b0}
-ig_{\eta_{b0}NN}\bar{N}\gamma_5N\eta_{b0}
\end{equation}
lead to a $q^2$-suppression factor at the rate level,
where $q$ is the momentum transfer of
the neutralino to the nucleon. An additional $q^2$ 
factor stems from the $A_{10}\chi\chi$ vertex, yielding altogether
a $(q^2/M_{N,\chi}^2)^2$ suppression factor in the scattering
cross section, where
$M_{N,\chi}$ stands for the target
and neutralino mass, respectively.

The effective coupling to the nucleon of 
either the mixed $A_1$ or $\eta_b$ state
via its hadronic component reads
\begin{equation}\label{eq:geff1}
g_{A_1NN}^{eff}=\sin{\alpha} \times g_{\eta_{b0}NN}\ \ ,\ \ 
g_{\eta_bNN}^{eff}=\cos{\alpha} \times g_{\eta_{b0}NN}\; .
\end{equation}
respectively, where the mixing angle $\alpha$  given by Eq.
(\ref{eq:alpha}).

In turn, the effective coupling of either 
the mixed $A_1$ or $\eta_b$ state to the
neutralino reads
\beq\label{eq:geff3}
g_{A_1\chi\chi}^{eff}=\cos{\alpha} \times g_{A_{10}\chi\chi}\ \ ,\ \ 
g_{\eta_b\chi\chi}^{eff}=-\sin{\alpha} \times g_{A_{10}\chi\chi} 
\eeq 
respectively, as
a result of its Higgs component.

\begin{figure}[ht!]
\begin{center}
\includegraphics*[width=0.9\linewidth]{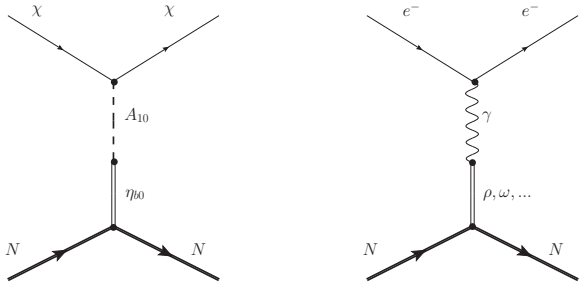}
\end{center}
\caption{Left: Contribution of a pseudoscalar mediator (either a
mixed $A_1$ or $\eta_b$ state) to the neutralino SD scattering cross
section. \ Right: Vector-meson-dominance model 
in the electron
scattering off nucleons \cite{PRINT-77-0549 (HARVARD)} 
where vector resonances ($\rho,
w ...)$ can couple to the exchanged photon.} \label{fig:2}
\end{figure}

Let us remark
that both $A_1$ and $\eta_b$  
are considered as physical states in our model, i.e. mass
eigenvalues of the full Hamiltonian without transitions
among their constituents apart from interactions with
external particles.
Note, however, that the WIMP scattering off nuclei
takes place at very low momentum transfer, so that
the pseudoscalar mediator in Fig.2 is quite off-shell.  
To the extent that the interaction amplitude is 
not too
sensitive to the energy difference between on-shell
and off-shell states, should our model make sense.

On the other hand, let us note that the $Z$-boson coupling to
the neutralino ($g_{Z\chi\chi}$)
should be of the same order as the
$A_{10}$ coupling, but    
the $A_{10}$-coupling to strange quarks would be tiny 
(especially for a dominantly singlet-like $A_{10}$) as
compared to the $Z$-boson coupling to nucleons ($g_{ZNN}$)
\cite{Das:2010ww}. 
However, the effective coupling (\ref{eq:geff1}) 
of mixed $A_1/\eta_b$ states to
the latter 
could be significantly enhanced 
via a non-perturbative effect.
Indeed, the experimental value 
${\cal B}[\eta_c \to p\bar{p}] \simeq 10^{-3}$ \cite{pdg}
turns out to be unexpectedly large in view of
the helicity
suppression resulting in a perturbative framework.
Even after including finite mass effects, a discrepancy of 
three or four orders of magnitude still remains 
with respect to theoretical calculations \cite{DFTT-04-92}.
We shall return later to this point of paramount importance
in our work, after
examining the spin-dependent $\chi$-nucleon cross section
mediated by a pseudoscalar Higgs versus a $Z$-boson.

\subsection{Pseudoscalar Higgs versus $Z$-boson exchange}

The $\chi$-nucleon 
scattering cross section with $A_1/\eta_b$ mixing
($\sigma_{SD}^{mix}$) can be estimated
from the ratio $r$$=\sigma_{SD}^{mix}/\sigma_{SD}^Z$,
where $\sigma_{SD}^Z$ denotes
the $Z$-exchange SD cross section, according to:
\begin{eqnarray}\label{eq:r}
r &\sim& \sin{}^2(2\alpha)
\biggl[\frac{g_{A_{10}\chi\chi}}{g_{Z\chi\chi}}\biggr]^2
\biggl[\frac{g_{\eta_{b0}NN}}{g_{ZNN}}\biggr]^2
\biggl[\frac{90 (\mathrm{GeV})}{m_{A_1}}\biggr]^4\ \times \nonumber \\
&& \biggl[\frac{q^2}{M_{N,\chi}^2}\biggr]^2 F^2[m_{A_{1}},m_{\eta_b},q^2]
\end{eqnarray}
where $F[m_{A_{1}},m_{\eta_b},q^2]$ takes into
account the interference effect due to the sign difference in 
Eqs.(\ref{eq:geff1}-\ref{eq:geff3}). At vanishing $q^2$ and the
range of masses considered in this work, one gets 
$F[m_{A_{1}},m_{\eta_b},0]$ $\simeq |m_{A_{1}}-m_{\eta_b}|/{m_{\eta_b}}$.

Actually a summation over all possible mixed states 
should be understood in (\ref{eq:r}), albeit with variable weight depending
on both their mixing strength
and the effective $\eta_{b0}(nS)$-nucleon coupling.

Let us first examine the ratio
\beq\label{ration}
\biggl[\frac{g_{A_{10}\chi\chi}}{g_{Z\chi\chi}}\biggr]^2
\simeq \frac{N_{11}^2N_{13}^2}{[N_{13}^2-N_{14}^2]^2} 
\frac{4\ g_1^2\cos^2{}\theta_W}{g_2^2}\ \cos{}^2\theta_A\; ,
\eeq
with $N_{1i}$ denoting the different (bino, higgsino ...) components
of the lightest neutralino, and  
the couplings $g_1$ and $g_2$ 
satisfy: $g_1^2/g_2^2 = \tan{}^2\theta_W$, where $\theta_W$ is the 
Weinberg angle.
Bounds from the invisible decay width of the
$Z$-boson 
imply that $|N_{13}^2-N_{14}^2|<0.11$.
Moreover, in our scan we always find $N_{11}$ close to unity
(at large $\tan{\beta}$ where also $N_{13}^2>>N_{14}^2$),
and $\cos{}^2(\theta_A) \in [10^{-2}-10^{-1}]$; hence we can write 
\beq\label{ration2}
\biggl[\frac{g_{A_{10}\chi\chi}}{g_{Z\chi\chi}}\biggr]^2
\simeq \frac{4N_{11}^2 
\sin{}^2\theta_W\cos{}^2\theta_A}{N_{13}^2} \sim
{\cal O}(10^{-1}-1)\; .
\eeq
\newline
Setting $m_{A_1}=10$ GeV, 
$\sin{}^2(2\alpha) \simeq 10^{-1}$
(typically expected from (\ref{eq:alpha}))
and $q^2$ $\simeq (100\ \mathrm{MeV})^2$
as reference values, we are led to
\beq\label{eq:ratiog}
\frac{g_{\eta_{b0}NN}}{g_{ZNN}} \gsim {\cal O}(10^3)\; ,
\eeq
so that the $A_1$/$\eta_b$  
exchange channel 
would become comparable numerically to the $Z$-exchange, i.e. 
$r={\cal O}(1)$. 

Naively one would expect the
above ratio to be of order
$\alpha_s/\sqrt{\alpha_{em}}$ (where $\alpha_s$ and 
$\alpha_{em}$ denote 
the strong interaction and electromagnetic coupling strengths, 
respectively) 
which hardly can yield such a large factor. 
Yet a big enhancement of (\ref{eq:ratiog})
could be plausible if 
a non-perturbative mechanism contributes to
the $\eta_{b0}$-nucleon coupling, as it likely happens
in the $\eta_c(1S)$ decay into a $p\bar{p}$ pair (though at 
quite larger $q^2$).

\subsection{Non-perturbative (instanton-induced) effects}

Indeed, an explanation of the large observed $\eta_c(1S) \to p\bar{p}$ 
decay rate seems to require a fundamental modification of the
perturbative approach to account for this decay mode.
Different proposals have been put forward in terms of 
a non-perturbative mechanism: mixing of the 
resonance and gluonium states \cite{hep-ph/9310344}, instanton effects 
\cite{Anselmino:1993bd}, 
intermediate meson loop contribution~\cite{Liu:2010um}
or higher Fock components \cite{Feldmann:2000hs}. 
Despite many uncertainties, it is conceivable that a
long-distance contribution also affects
$\eta_{b0}(nS)$ resonances.

Let us focus hereafter on instanton effects, which play
a fundamental role in understanding the QCD vacuum and many other
topics related to hadronic physics,  
especially concerning light hadrons (see e.g. \cite{Schafer:1996wv}
for a review). Nevertheless, instanton effects can still be
relevant in heavy quark systems, e.g. for the non-perturbative
gluon condensate in charmonium \cite{Novikov:1977dq}. 

In particular, the authors of 
Ref.~\cite{Anselmino:1992wi} studied 
the instanton contribution to
non-perturbative chiral symmetry breaking 
in proton-proton scattering at high energy. The same authors
later considered this interaction contributing
to the decay of the  
$\eta_c$ resonance into a $p\bar{p}$ pair 
\cite{Anselmino:1993bd,Zetocha:2002as}. 
The idea is that the meson resonance
annihilates into two gluons (perturbative part)
that are absorbed by instantons,
which couple to a baryon pair (non-perturbative part). 

In the following we envisage whether such an 
instanton-induced interaction could still affect
the $\eta_b$ resonance coupling to baryons for
a momentum transfer from $q=m_{\eta_b(1S)}$ 
down to $q= 100$ MeV. If so,  
the WIMP scattering off nuclei 
might then bear an unexpected
resemblance to $p$-$p$ elastic scattering (at small $-t$).

Instanton effects are usually
assumed to depend linearly on the instanton
density given by \cite{Anselmino:1993bd,'tHooft:1976fv}: 
\beq\label{eq:exp} 
\frac{dn(\rho)}{d\rho} \sim \frac{1}{\rho^5}\ [\alpha_s(\rho^{-1})]^{-6}\
\exp{\biggl[-\frac{2\pi}{\alpha_s(\rho^{-1})}\biggr]}\; ,
\eeq
where $\rho$ denotes the instanton size (such that $\rho \lesssim 1/q$).
Quite in general, instanton effects are expected to be weighted by
the Euclidean instanton action of the exponential factor in
Eq.(\ref{eq:exp}), becoming
more relevant at smaller $q^2$ (hence larger $\alpha_s$).

Now, in order to obtain an estimate of the 
$\eta_b(1S) \to p\bar{p}$ decay rate within this framework, 
we first rescale the perturbative part of the calculation
in \cite{Anselmino:1993bd}
to the bottomonium system according to:
\beq
\frac{|R_{\eta_b}(0)|^2}{|R_{\eta_c(0)}|^2}\ \times\  
\frac{\alpha_s^2(m_{\eta_b})}{\alpha_s^2(m_{\eta_c})}\ \times\ 
\frac{m_{\eta_c}^2}{m_{\eta_b}^2}\ 
\approx\ 0.3\; ,
\eeq 
where the wave function of the spin-singlet state 
can be approximated by the corresponding spin-triplet one
\cite{Eichten:1995ch}.

Turning now to the non-perturbative part of the calculation, 
we will assume (as usual)
that the instanton-induced coupling to nucleons depends
on the number of instantons $n(1/q)$ relevant in the process,  
obtained by integration of the instanton density 
(\ref{eq:exp}) over $\rho \leq 1/q$. Note
that the resulting value should be controlled by  
the exponential dependence on $\rho$ around the inverse mass
of the heavy resonance. Varying the momentum transfer to baryons, 
from $q=m_{\eta_c}$ to $q=m_{\eta_b}$, 
$n(1/q)$ decreases roughly by two orders of magnitude, hence
lowering the rate (which depends on $n(1/q)^2$) by about
four orders of magnitude.

Even though phase space favours the $\eta_b \to p\bar{p}$ decay rate
with respect to the $\eta_c$ by a factor $\approx 1.3$, 
we finally conclude, taking into account both 
perturbative and non-perturbative parts, that the partial 
width $\Gamma[\eta_b \to p\bar{p}]$ 
induced by instanton effects should
be about four orders of magnitude (though with large uncertainties) 
smaller than $\Gamma[\eta_c \to p\bar{p}]$. The corresponding
BF can be obtained making use of the central value for the
$\eta_b(1S)$ full width $\Gamma_{\eta_b(1S)}= 12.4$ MeV, recently 
found by Belle \cite{Collaboration:2011chb}. 

In sum, our order-of-magnitude
estimate based on instanton-induced interaction reads:
\beq\label{eq:BFeta}
{\cal B}[\eta_b \to p\bar{p}] \approx 10^{-7}-10^{-6}  
\eeq

Yet the $\eta_b(1S)$ can decay into a $p\bar{p}$ pair
at an observable rate at B (Super) factories \cite{Bona:2007qt}.  
The experimental 
determination of (\ref{eq:BFeta}) thus becomes
relevant to uncover possible non-perturbative effects
(the present upper limit being 
$ 5 \times 10^{-4}$ \cite{pdg}) associated to
the $\eta_b$ state, permitting a reliable comparison with perturbative QCD
predictions because of the heavier bottom mass. 

On the other hand, 
as already commented, instanton-induced effects should become
quite more important at low momentum transfer because
of the exponential in Eq.(\ref{eq:exp}). Thus, 
a sizable non-perturbative effect in $\eta_b$ decays
into baryons occurring at $q^2 \simeq m_{\eta_b}^2$
should be enhanced (actually less
suppressed) at the much
lower energy scale set by the small momentum transfer
in WIMP scattering off nuclei, $q$ $\simeq 100\ \mathrm{MeV}$.
However, extrapolation to such low $q$ value from the
$\eta_c$ or $\eta_b$ mass
requires a further and detailed examination \cite{new}.

Furthermore, our previous caveat concerning the off-shellness of
the pseudoscalar mediator of Fig.2 is in order as indeed
$q^2 << m_{\eta_b}^2$. In particular, its hadronic 
component could display in the nucleon-nucleon vertex
a different behaviour than an on-shell $\eta_b$ resonance. 
Similarly, in a non-covariant language, one can invoke 
the time-energy uncertainty principle implying
that the time for a virtual $b\bar{b}$ pair is likely too
short for the $\eta_b$ bound state to be formed. Therefore the
above calculation of the perturbative part 
cannot not straightforwardly be applied. 
Nonetheless, quantum numbers of the virtual $b\bar{b}$ pair
should still correspond to
a pseudoscalar state thereby permitting two
gluons to be emitted, ultimately leading to 
instanton-induced (spin-dependent) effective interaction 
of WIMPs with nuclei. 

Conversely, such a mechanism
should not significantly affect neutralino annihilation 
into SM particles
via a $s$-channel exchange of a 
CP-odd Higgs boson, for the energy scale 
would be again of order ${\cal O}(10)$ GeV, thereby 
avoiding an extra tension with
indirect detection limits, such as cosmic-ray
antiprotons \cite{Bottino:2005xy,Lavalle:2010yw}.

\section{SUMMARY}

In this paper,  
we have considered a particular scenario within the NMSSM
with both a light neutralino and a light CP-odd Higgs
boson, the latter sizably
mixing with pseudoscalar $\eta_b$ resonances.  
Implicit in our work is the idea that 
non-perturbative effects (e.g. instanton-induced interaction)
may lead to a non-negligible pseudoscalar contribution to the 
$\chi$-nucleus scattering, thereby introducing a momentum-dependent
form factor in the cross section 
which might be helpful (see e.g. Ref.~\cite{Feldstein:2009tr}) to 
interpret the results of direct DM search experiments, 
with variable sensitivity along the
nuclear recoil energy range. 

To conclude we stress that an accurate experimental
test of lepton universality in $\Upsilon$ decays,
the discovery of the $\eta_b(2S)$ resonance 
together with the measurements of
${\cal B}[\eta_b(nS) \to p\bar{p}]$ at a (Super) B factory
\cite{Bona:2007qt} could be relevant for a better understanding of
DM searches and related astrophysical questions.

\subsection*{ACKNOWLEDGMENTS}

This work was supported by research grants FPA2011-23596
and GVPROMETEO2010-056.
We thank M.~Baker, A.~Djouadi, F.~Domingo, S.~Eidelman, 
N.~Kochelev and J.~Nieves
for many useful comments.


\begin{thebibliography}{99}



\bibitem{Bernabei:2010mq} R.~Bernabei {\it et al.},
  Eur.\ Phys.\ J.\  {\bf C67}, 39-49 (2010).

\bibitem{Aalseth:2011wp} C.~E.~Aalseth {\it et al.},
  Phys.\ Rev.\ Lett.\  {\bf 107}, 141301 (2011).

\bibitem{Angloher:2011uu}
  G.~Angloher {\it et al.},
  [arXiv:1109.0702 [astro-ph.CO]].

\bibitem{arXiv:1104.3088} 
  J.~Angle {\it et al.} [XENON10 Collaboration],
  Phys.\ Rev.\ Lett.\ \ {\bf 107}, 051301 (2011).

\bibitem{arXiv:1104.2549} 
  E.~Aprile {\it et al.} [XENON100 Collaboration],
  Phys.\ Rev.\ Lett.\ \ {\bf 107}, 131302 (2011).

\bibitem{Gunion:2005rw} J.~F.~Gunion, D.~Hooper, B.~McElrath,
  Phys.\ Rev.\  {\bf D73}, 015011 (2006).
  [hep-ph/0509024].


\bibitem{Das:2010ww} D.~Das, U.~Ellwanger,
  JHEP {\bf 1009}, 085 (2010).

\bibitem{Gunion:2010dy}
  J.~F.~Gunion, A.~V.~Belikov, D.~Hooper,
  arXiv:1009.2555.


\bibitem{arXiv:0910.1785} 
  U.~Ellwanger, C.~Hugonie and A.~M.~Teixeira,
  Phys.\ Rept.\ \ {\bf 496}, 1 (2010)
  [arXiv:0910.1785 [hep-ph]].

\bibitem{Domingo:2008rr} F.~Domingo {\it et al.},
  JHEP {\bf 0901}, 061 (2009).


\bibitem{Ananthanarayan:1996zv}
  B.~Ananthanarayan and P.~N.~Pandita,
  Int.\ J.\ Mod.\ Phys.\  A {\bf 12} (1997) 2321
  [arXiv:hep-ph/9601372].


\bibitem{Fullana:2007uq} E.~Fullana and M.~A.~Sanchis-Lozano,
  Phys.\ Lett.\  B {\bf 653} (2007) 67
  [arXiv:hep-ph/0702190].

\bibitem{Domingo:2009tb} F.~Domingo, U.~Ellwanger and M.~A.~Sanchis-Lozano,
  {\em Phys.\ Rev.\ Lett.}\  {\bf 103}, 111802 (2009).

\bibitem{Drees:1989du}
  M.~Drees and K.~i.~Hikasa,
  Phys.\ Rev.\  D {\bf 41} (1990) 1547.


\bibitem{SanchisLozano:2010ij} M.~A.~Sanchis-Lozano,
  arXiv:1003.0312 [hep-ph].

\bibitem{hep-ph/0612031} R.~Dermisek, J.~F.~Gunion and B.~McElrath,
  Phys.\ Rev.\ D\ {\bf 76}, 051105  (2007) [hep-ph/0612031].


\bibitem{Sanchis-Lozano:2003} M.~A.~Sanchis-Lozano,
  Int.\ J.\ Mod.\ Phys.\ A \textbf{19}, 2183 (2004) [arXiv:hep-ph/0307313].

\bibitem{SanchisLozano:2006gx} 
  M.~-A.~Sanchis-Lozano,
  J.\ Phys.\ Soc.\ Jap.\  {\bf 76}, 044101 (2007).

\bibitem{Roney} M.~Roney, private communication.

\bibitem{pdg} K. Nakamura {\em et al.} (Particle Data Group),
  J.\ Phys.\ G {\bf 37}, 075021 (2010)

\bibitem{Kniehl:2003ap} B.~A.~Kniehl {\it et al.},
  Phys.\ Rev.\ Lett.\  {\bf 92}, 242001 (2004).



\bibitem{Collaboration:2011chb} 
  B.~Collaboration,
  arXiv:1110.3934 [hep-ex].


\bibitem{Meinel:2010pv} S.~Meinel,
  Phys.\ Rev.\  D {\bf 82}, 114502 (2010).

\bibitem{nmssmtools}
  U.~Ellwanger and C.~Hugonie,
  Comput.\ Phys.\ Commun.\  {\bf 177}, 399 (2007).

\bibitem{Domingo:2010am} F.~Domingo,
  JHEP {\bf 1104}, 016 (2011).


\bibitem{2008st} B.~Aubert {\it et al.}  [BaBar Collaboration],
  arXiv:0808.0017.

\bibitem{astro-ph/0408346} C.~Savage, P.~Gondolo and K.~Freese,
  Phys.\ Rev.\ D\ {\bf 70}, 123513  (2004)
  [astro-ph/0408346].

\bibitem{Chang:2009yt} S.~Chang, A.~Pierce, N.~Weiner,
  JCAP {\bf 1001}, 006 (2010). 

\bibitem{Freytsis:2010ne} M.~Freytsis, Z.~Ligeti,
  Phys.\ Rev.\  {\bf D83}, 115009 (2011).

\bibitem{PRINT-77-0549 (HARVARD)} T.~H.~Bauer, {\it et al},
  Rev.\ Mod.\ Phys.\ \ {\bf 50}, 261  (1978).


\bibitem{Draper:2010ew} 
  P.~Draper {\it et al},
  Phys.\ Rev.\ Lett.\  {\bf 106}, 121805 (2011)
  [arXiv:1009.3963 [hep-ph]].


\bibitem{DFTT-04-92} M.~Anselmino, R.~Cancelliere and F.~Murgia,
  Phys.\ Rev.\ D\ {\bf 46}, 5049  (1992).

\bibitem{hep-ph/9310344} M.~Anselmino, M.~Genovese and D.~E.~Kharzeev,
  Phys.\ Rev.\ D\ {\bf 50}, 595  (1994)
  [hep-ph/9310344].

\bibitem{Anselmino:1993bd} 
  M.~Anselmino and S.~Forte,
  Phys.\ Lett.\ B {\bf 323}, 71 (1994)
  [hep-ph/9311365].

\bibitem{Liu:2010um} X.~-H.~Liu, Q.~Zhao,
  J.\ Phys.\ G {\bf G38 }, 035007 (2011).



\bibitem{Feldmann:2000hs} 
  T.~Feldmann and P.~Kroll,
  Phys.\ Rev.\ D {\bf 62}, 074006 (2000)
  [hep-ph/0003096].

\bibitem{Schafer:1996wv} 
  T.~Schafer and E.~V.~Shuryak,
  Rev.\ Mod.\ Phys.\  {\bf 70}, 323 (1998)
  [hep-ph/9610451].

\bibitem{Novikov:1977dq} 
  V.~A.~Novikov, L.~B.~Okun, M.~A.~Shifman, A.~I.~Vainshtein, M.~B.~Voloshin and V.~I.~Zakharov,
  Phys.\ Rept.\  {\bf 41}, 1 (1978).

\bibitem{Anselmino:1992wi} 
  M.~Anselmino and S.~Forte,
  Phys.\ Rev.\ Lett.\  {\bf 71}, 223 (1993)
  [hep-ph/9211221].

\bibitem{Zetocha:2002as} 
  V.~Zetocha and T.~Schafer,
  Phys.\ Rev.\ D {\bf 67}, 114003 (2003)
  [hep-ph/0212125].


\bibitem{'tHooft:1976fv} 
  G.~'t Hooft,
  Phys.\ Rev.\ D {\bf 14}, 3432 (1976)
  [Erratum-ibid.\ D {\bf 18}, 2199 (1978)].


\bibitem{Bona:2007qt} M.~Bona {\it et al.},
  arXiv:0709.0451 [hep-ex].


\bibitem{new} Work in progress.

\bibitem{Bottino:2005xy} 
  A.~Bottino, F.~Donato, N.~Fornengo and P.~Salati,
  Phys.\ Rev.\ D {\bf 72}, 083518 (2005)
  [hep-ph/0507086].


\bibitem{Eichten:1995ch} 
  E.~J.~Eichten and C.~Quigg,
  Phys.\ Rev.\ D {\bf 52}, 1726 (1995)
  [hep-ph/9503356].

\cite{Lavalle:2010yw}
\bibitem{Lavalle:2010yw} 
  J.~Lavalle,
  Phys.\ Rev.\ D {\bf 82}, 081302 (2010)
  [arXiv:1007.5253 [astro-ph.HE]].



\bibitem{Feldstein:2009tr} 
  B.~Feldstein, A.~L.~Fitzpatrick and E.~Katz,
  JCAP {\bf 1001}, 020 (2010)
  [arXiv:0908.2991 [hep-ph]].

\end{thebibliography}
\end{document}